\documentclass[prd,aps,twocolumn,showpacs]{revtex4} 
\usepackage{amssymb}
\usepackage{graphicx}

\begin{document}

\draft
\title{ Brane Induced Gravity in the Curved Bulk}  
\author{			Keiichi Akama}
\address{	Department of Physics, Saitama Medical University,
 			 Saitama, 350-0495, Japan}
\author{			Takashi Hattori}
\address{	Department of Physics, Kanagawa Dental College,
 			 Yokosuka, 238-8580, Japan}
\date{\today}

\begin{abstract}
Starting with the Nambu-Goto action of the braneworld embedded in a curved bulk,
	we derive the precise expressions 
	for the quantum induced effects due to small fluctuations of the brane.
To define the brane fluctuations invariantly,
	we introduce the Riemannian coordinate system for 
	the subspace normal to the braneworld. 
It will turn out that we can systematically incorporate the effects of bulk curvature, 
	and that the induced effects depend
	on the extrinsic curvature and the normal-connection gauge field
	as well as on bulk curvature components at the brane.
\end{abstract}





\maketitle

\section{Introduction}

Recently, the ideas of braneworld and brane induced gravity
	\cite{Fronsdal}--\cite{Akama:2013tua}
	attracts much attention in wide areas of physics
	such as particle physics, field theory, superstring theory and cosmology.
In the previous paper \cite{Akama:2013tua}, 
	we inquired their dynamical foundations
	and established the precise formulations
	for a rather limited setup of the flat bulk.
We fixed imperfections in the naive old formalism.
For example, the induced gravity terms are proportional to
	the number of the extra dimensions, but not to that of whole spacetime.
We showed that the induced effects depend also on the extrinsic curvature
	and the normal-connection gauge field,
	and derived the precise expressions.
Delicate situations for the cosmological-constant problem were discussed.
Things were much simplified due to the flatness of bulk.
In this paper, we extend the formulation to the fully general case of the curved bulk.
It will turn out that we can incorporate the bulk curvature effects 
	in a systematic way, as far as the small fluctuations are concerned.
The induced effects depend also on the bulk curvature components at the brane,
	in addition to the extrinsic curvature and the normal-connection gauge field.

General relativity is based on the premises  
that the spacetime is curved affected by matter 
	according to the Einstein equation, 
	and that the objects move along the spacetime geodesics.
The gravitations are apparent phenomena 
	of the inertial motions in the curved spacetime. 
This successfully explains 
	why the motion in the ``gravitational" field is universal, 
	i.e. blind to the object properties (mass, charge etc.), 
	and why the Newton's law of gravitation holds
	in daily familiar gravitational phenomena.
The general relativity is supported by further precise observations such as
	perihelion precessions of planets, light bending due to massive stars etc..
It raises, however, another fundamental question 
	{why the spacetime is curved so as the Einstein equation indicates.}
It is supposed by Sakharov in 1967 that 
	the quantum effects of the spacetime itself may provide 
	the origin of the gravity with the Einstein equation \cite{Sakharov}. 
Field theoretical formulations were developed 
	to realize the Einstein gravity as quantum effects of the matters
	\cite{indg}--\cite{indg3}.
In these theories, the gravitational field, 
	i. e. spacetime metric appears as a composite of the matters.

On the other hand, the composite pictures of matters achieved 
	great successes in wide areas of physics, including
	molecules, atoms, atomic nuclei, hadrons and various quasi-particles
	in quantum systems. 
Field theoretical treatments of particle compositeness 
	have been explored extensively.
According to Feynman rules, scattering amplitudes have a pole due to a particle
	in its energy variable at its mass.  
If a pole appears in some energy channel through some mechanism
	without a corresponding particle field in the original setup, 
	it indicates existence  of a composite particle formed by the mechanism.
For example, contact interactions of fermions 
	are known to give rise to a composite pole through chain diagrams,
	when we fix the momentum cut off at a large but finite level
	\cite{NJL}. 
Some degrees of freedom at the short distances are 
	converted into those of the collective modes.
Such a composite is also successfully described by introducing an auxiliary field. 
If we eliminate the auxiliary field by using the constraint
	from the Euler equation in the original setup,
	we have a system described only with constituents.
Its kinetic term is supplied through quantum loop diagrams.
The methods are applied to, for example, superconductivity, 
	models of hadrons, induced gauge theories \cite{Bjorken}, and
	models of composite quarks, leptons, gauge bosons, and Higgs bosons
	\cite{TCA}.
In renormalization theories (with large but finite momentum cut off), 
	vanishing of the wave-function renormalization constant of some field
	implies absence of the kinetic term of the field in the original setup, 
	despite its presence after renormalization. 
Then, the field is interpreted as a composite \cite{compcond}.  
Vanishing of renormalization constants 	is called as compositeness condition.
These methods have been extensively studied \cite{compcond2} and
	widely applied in condensed-matter, nuclear, and particle physics.

The field theoretical formulations of the induced gravity \cite{indg}
	is developed in the context of the unified composite model \cite{TCA}.
In the original setup, the system is described only with matter fields
	with the assumption of general coordinate invariance.
Or, equivalently, it is described with metric field
	which lacks the kinetic term, and is taken as an auxiliary field
	without independent degree of freedom.
The kinetic term of the metric, i.\ e.\ the Einstein-Hilbert action,
	is induced via quantum loop diagrams, 
	and the metric acquires the independent degrees of freedom.
Thus, the metric is interpreted as the composite of the matters,
	and the origin of the gravitational phenomena is
	traced back to the quantum nature of the matters.
Unfortunately, we can only partially apply the renormalization theory arguments
	because the general relativity is not renormalizable.

The simplest model of matters with general-coordinate invariance 
	is that of scalar fields with the Nambu-Goto action 
 	\cite{Akama78}.
Then, the scalar fields can be interpreted 
	as the position coordinate of our spcetime 
	in a spacetime with higher dimensions.
Note that the spacetime itself is taken as a matter,
	which may induce quantum effects including gravity.
This interpretation lead us to the ideas of the braneworld
	and the brane induced gravity \cite{Akama82},
	\cite{Pavsic}, \cite{Akama87}--\cite{Pavsic:1994rq}. 
These ideas
	have been studied extensively in these three decades. 
Physical models were constructed with topological defects 
	in higher dimensional spacetime \cite{Akama82}--\cite{Akama88b},
	and they were realized as ``D-branes" in the superstring theory 
	\cite{DaiLeighPolchinski}--\cite{HoravaWitten}.
They were applied to the hierarchy problem 
	with large extra dimensions, or with warped extra dimensions 
	\cite{Antoniadis}, \cite{ADD}--\cite{RS}.
It was argued that the brane induced gravity would imply
	various interesting consequencs 
	\cite{GregoryRubakovSibiryakov}--\cite{DvaliGabadadze}.
The ideas have been studied in wide areas including
	basic formalism, \cite{Th1}--\cite{Th4},
	brane induced gravity \cite{AH}--\cite{Krause:2008sj},
	particle physics phenenomenology \cite{Ida:2002ez}--\cite{SarrazinPetit},
	and cosmology \cite{KantiKoganOlivePospelov}--\cite{Jardim:2011gg}.

In this paper, 
	we explore a precise formalism to derive the expressions 
	for the quantum induced effects on the brane
	embedded in a curved bulk.
For definiteness, we follow 
	the simplest model of the braneworld 
	with the Nambu-Goto dynamics. 
We do not specify dynamics of the bulk gravity,
	since it is irrelevant to the short distance effects at the brane.
We treat the bulk curvature only as a given external field.
To define the brane fluctuations invariantly,
	we introduce a geodesic measure 
	(Riemannian coordinate) for 
	the normal subspace to the braneworld. 
Then, we work out the quantum effects for the small fluctuation
	using the methods developed in the previous paper.
It will turn out that we can systematically incorporate the effects of bulk curvature, 
	and that the induced effects depend also 
	on the bulk curvature components at the brane,
	as well as on the extrinsic curvature and the normal-connection gauge field.
The plan of this paper is as follows.
First, we define the model (Sec.\ \ref{model}),    
	and then 
	we derive the quantum effects 
	(Secs.\ \ref{fluctuations}--\ref{calculation}). 
We define the brane fluctuations (Sec.\ \ref{fluctuations}), 
	formulate the quantum effects (Sec.\ \ref{quantum}), 
	specify the method to regularize the divergences (Sec.\ \ref{cutoff}), 
	classify the possible induced terms according to symmetries
	 (Sec.\ \ref{classification}), 
	and calculate them via Feynman diagram method (Sec.\ \ref{calculation}).
The final section (Sec.\ \ref{conclusion}) is devoted to conclusion and discussions.
The cosmological terms are fine-tuned, and
	the Einstein like gravity and other terms are induced .

\section{The Model \label{model}}

We consider a quantum theoretical braneworld described 
	by the  Nambu-Goto Lagrangian \cite{NambuGoto}. 
We will see the quantum effects of the brane fluctuations
	give rise to effective braneworld gravity \cite{Akama82},
	\cite{Pavsic}, \cite{Akama87}--\cite{Pavsic:1994rq}. 
Let $X^I(x^\mu)$ $(I=0,1,\cdots,D-1)$ 
	be the position of our three-brane in the $D$ dimensional spacetime (bulk), 
	parameterized by the brane coordinate $x^\mu$ $(\mu=0,1,2,3)$,
	where $I=0$ and $\mu=0$ indicate the time components.
Let $G^{IJ}(X^K)$ be the bulk metric tensor at the bulk point $X^K$.
This is taken to obey some bulk gravity theory.
Then we consider a braneworld with dynamics given 
	by the Nambu-Goto Lagrangian (density): 
\begin{eqnarray}
	{\cal L}_{\rm br}=-\lambda \sqrt{
	-\det_{\mu\nu} \left(
	\frac{\partial X^I}{\partial x _\mu}
	\frac{\partial X^J}{\partial x _\nu}
	G_{IJ}(X^K)
	\right)},
  \label{NG}
\end{eqnarray}
	where $\lambda$ is a constant.
Or we write it as
\begin{eqnarray}
	{\cal L}_{\rm br}=-\lambda \sqrt{- g^{[X]} }  \label{NGab}
\end{eqnarray}
with abbreviations $ g^{[X]} =\det g^{[X]}_{\mu\nu}$, and
\begin{eqnarray}
	{ g }^{[X]} _{\mu\nu} =X^I_{\ ,\mu} X^J_{\ ,\nu} G_{IJ}(X^K), 
  \label{gmunuX}
\end{eqnarray}
where (and hereafter) indices following a comma (,) indicate differentiation
	with respect to the corresponding coordinate component, 
	and ${[X]}$ is attached to remind 
	that they are abbreviations for expressions written in terms of $X^I$.
The tensor $g^{[X]}_{\mu\nu}$ is the induced metric on the brane
	with (\ref{gmunuX}). 
We assume that $X^I$ appears nowhere other than in ${\cal L}_{\rm br}$ 
	in the total Lagrangian ${\cal L}_{\rm tot}$ including the bulk Lagrangian.
The equation of motion from (\ref{NG}) is given by
\begin{eqnarray}
	g ^{[X]} {}^{\mu\nu} X ^{[X]}_{;\mu\nu}{} ^I =0,
  \label{NGEM}
\end{eqnarray}
	where $ X ^{[X]}_{;\mu\nu} {}^I $ is the double covariant derivative
	with respect to both of the general coordinate transformations 
	on the brane and to those in the bulk:
\begin{eqnarray} 
	X ^{[X]}_{;\mu\nu} {}^I = X^I_{\ ,\mu\nu}
	-X^I_{\ ,\lambda} \gamma^{[X]}{}^\lambda_{\mu\nu}
	+X^J_{\ ,\mu} X^K_{\ ,\nu}\Gamma^I_{JK}
\label{X[X];munu}
\end{eqnarray}
	with the affine connections on the brane and bulk
\begin{eqnarray}
	\gamma^{[X]}{}^\lambda_{\mu\nu}
	&=&\frac{1}{2} g^{[X]}{} ^{\lambda \rho}\left(
	g ^{[X]}_{\rho\mu,\nu}+ g^{[X]}_{\rho\nu,\mu}- g^{[X]} _{\mu\nu,\rho}
	\right),
\\
	\Gamma^I_{JK}
	&=&\frac{1}{2} G^{IL}\left(
	G_{LJ,K}+ G_{LK,J}- G_{JK,L}
	\right),
\label{GammaBulk}
\end{eqnarray}
	respectively.
The system is invariant under the general coordinate transformation 
	of the bulk and the brane separately.
Under these symmetries, 
	we can also have terms dependent on the curvature tensor 
	written with ${ g }^{[X]} _{\mu\nu}$.
They would, however, be suppressed for small curvatures
	as our exiting spacetime.
Therefore we concentrate on the case where the Lagrangian 
	is dominated by the simplest form (\ref{NG}).
We expect that this gives a good approximation at low curvature limit 
	in many dynamical models of the braneworld 
	(e.g. topological defects \cite{Akama82}--\cite{Akama88b}, 
	spacetime singularities \cite{Antoniadis}, \cite{ADD}--\cite{RS}, 
	D-branes \cite{DaiLeighPolchinski}--\cite{HoravaWitten}, etc.).
It is remarkable that, as we shall see below, this simple model exhibits 
	brane gravity and gauge theory like structure through the quantum effects.

For convenience of quantum treatments, 
	we consider the following equivalent Lagrangian 
	to (\ref{NG}):
\begin{eqnarray}
	{\cal L}'_{\rm br}=-\frac{\lambda}{2} \sqrt{-g}\left[
	g^{\mu\nu}X^I_{\ ,\mu} X^J_{\ ,\nu}G_{IJ}(X^K)-2
	\right]
  \label{NG'}
\end{eqnarray}
where $g_{\mu\nu}$ is an auxiliary field, $g=\det g_{\mu\nu}$, and
	$g^{\mu\nu}$ is the inverse matrix of $g_{\mu\nu}$.
Note that $g_{\mu\nu}$, unlike $ g^{[X]}_{\mu\nu}$ above,
	is treated as a field independent of $X^I$.
Then the Euler Lagrange equations with respect to $X^I$ and $g_{\mu\nu}$ 
	are given by 
\begin{eqnarray}&&\hskip-20pt
	g^{\mu\nu} X _{;\mu\nu} ^I =0,
  \label{NG'EM}
\\&&\hskip-20pt 
	g_{\mu\nu}=X^I_{\ ,\mu} X^J_{\ ,\nu}G_{IJ}(X^K),
  \label{gmunu}
\end{eqnarray}
	respectively, where the covariant derivative 
\begin{eqnarray}
	X _{;\mu\nu}{} ^I = X^I_{\ ,\mu\nu}
	-X^I_{\ ,\lambda} \gamma^\lambda_{\ \mu\nu}
	+X^J_{\ ,\mu} X^K_{\ ,\nu}\Gamma^I_{\ JK}
\label{X;munu}
\end{eqnarray}
is written in terms of the brane affine connection 
\begin{eqnarray}
	\gamma^\lambda_{\ \mu\nu}
	&=&\frac{1}{2} g ^{\lambda \rho}\left(
	g _{\rho\mu,\nu}+ g _{\rho\nu,\mu}- g _{\mu\nu,\rho}
	\right)
\label{gamma}
\end{eqnarray}
	with respect to the auxiliary field $g_{\mu\nu}$.
Now $g_{\mu\nu}$ in (\ref{NG'EM}) is independent of $X^I$,
	and, instead, we have an extra equation (\ref{gmunu}),
	which guarantees that $g_{\mu\nu}$ is the induced metric.
If we substitute (\ref{gmunu}) into (\ref{NG'EM}), 
	we obtain the same equation as (\ref{NGEM}).
Thus the systems with the Lagrangians ${\cal L}_{\rm br}$ 
	and ${\cal L}'_{\rm br}$ coincide.
Furthermore the argument that their Dirac bracket algebrae coincide 
	\cite{Akama79} indicates their quantum theoretical equivalence.
We proceed hereafter based on the Lagrangian ${\cal L}'_{\rm br}$
	instead of ${\cal L}_{\rm br}$.

\section{Brane Fluctuations\label{fluctuations}}

In order to extract the quantum effects of ${\cal L}'_{\rm br}$,
	we deploy a semi-classical method, 
	where we consider those due to small fluctuations of the brane 
	around some classical solution (say $Y^I(x^\mu)$) for $X^I(x^\mu)$
	of the equation of motion (\ref{NGEM}) \cite{Watanabe}.
Namely, the solution $Y^I(x^\mu)$ obeys the classical equation
\begin{eqnarray}&&
	g^{\mu\nu} Y _{;\mu\nu}^I =0,
  \label{NG'EMY}
\\&&
	g_{\mu\nu}=Y^I_{\ ,\mu} Y^J_{\ ,\nu}G_{IJ}(Y^K).
  \label{gmunuY}
\end{eqnarray}
In quantum treatment, 
	$X^I$ itself in the Lagrangian ${\cal L}'_{\rm br}$ does not necessarily
	obey the equation of motion (\ref{NGEM}), 
	and may fluctuate from $Y^I(x^\mu)$. 
{Among the fluctuations, 
	only those transverse to the brane are physically meaningful},
	because those along the brane remain within the brane
	and cause no real fluctuations of the brane.   
They are absorbed by general coordinate transformations.
In order to describe them, we choose $D-4$ independent normal vectors 
	$n_j{}^I (x^\mu)$ $(j=4,\cdots,D-1)$ 
	at each point on the brane with the orthogonality condition
\begin{eqnarray}&&
	n_j{}^I Y^J_{\ ,\nu} G _{IJ}(Y^K) =0.
  \label{nY}
\end{eqnarray}
We can arbitrarily choose a orthonormal system with
\begin{eqnarray}&&
	n_i{}^I n_j{}^J G _{IJ}(Y^K) =\eta_{ij}\equiv-\delta_{ij},
  \label{nn}
\end{eqnarray}
where $\delta_{ij}$ is the Kronecker delta.
	Then, we have the completeness relation of the vectors, 
\begin{eqnarray}&&
	Y^I_{\ ,\mu}Y^J_{\ ,\nu} g^{\mu\nu}+n_i^{\ I} n_j^{\ J} \eta^{ij}
	= G ^{IJ}(Y^K).
  \label{YY+nn=G}                                                        
\end{eqnarray}
Throughout this paper,
	Latin capital sufficies $I,J,K,\cdots$
	indicate bulk coordinate indices running over 
	the range $0,1,2,\cdots,D-1$, 
	Greek lower case sufficies $\mu,\nu,\lambda,\cdots$
	indicate brane coordinate indices running over 
	the range $0,1,2,3$, and 
	Latin lower case sufficies $i,j,k,\cdots$
	indicate extra-dimensional coordinate indices running over 
	the range $4,5,\cdots,D-1$. 
Bulk coordinate indices $I,J,\cdots(=0,\cdots,D-1) $ are raised and lowered
	by the metric tensors $G_{IJ}$ and $G^{IJ}$.
We can read off from (\ref{NG'}) and (\ref{gmunu})
	that the auxiliary field $g_{\mu\nu}$ 
	plays the role the metric tensor on the brane.
On the brane, we raise and lower 
	the brane coordinate indices $\mu,\nu,\cdots(=0,\cdots,3) $ 
	by $g_{\mu\nu}$ and $g^{\mu\nu}$ 
	(but not by $g^{[X]}_{\mu\nu}$ and $g^{[X] \mu\nu}$),
	and 
	the normal space indices $i,j,k,\cdots(=4,\cdots,D-1) $ 
	by  $\eta_{ij}$ and $\eta^{ij}$.

In the previous paper, we assumed that the bulk is flat,
	and we defined the fluctuation measure $\phi^i(x^\mu)$ with
\begin{eqnarray}
	X^I=Y^I+\phi^i n_i{}^I.
  \label{X=Y+prev}
\end{eqnarray}
In this paper, we want to consider general cases where the bulk is also curved.
Then, the definition of the fluctuation measure 
	$\phi^i(x^\mu)$ with (\ref{X=Y+prev}) is inappropriate,
	because it relies on the coordinate system of $X^I$,
	and lacks the general-coordinate invariance of the bulk.
We define the invariant measure $\varphi^i(x^\mu) $ 
	for quantum fluctuations of the brane as follows.  
Suppose that the geodesic curve in the direction 
	of a normal unit vector $n^I(x^\mu)$
	of the solution brane $Y^I$ 
	hits the fluctuated brane $X^I$ at a distance $s$.
Then, $\varphi^i$ is the coefficients of the expansion of $sn^I$
	in terms of $ n_i{}^I $: 
\begin{eqnarray}&&
	sn^I=\varphi^i n_i{}^I. 
  \label{sn}
\end{eqnarray}
The position $X^I$ of the fluctuated brane is given by
\begin{eqnarray}
	X^I=Y^I+\varphi^i n_i{}^I
	-\frac{1}{2}\Gamma^I_{JK}\varphi^i n_i{}^J\varphi^j n_j{}^K+\cdots, &&
  \label{X=Y+}
\end{eqnarray}
where $\cdots$ stands for terms of O($ (\varphi^i)^3$) and the higher,
	and recursively given by solving the geodesic equation
	\cite{Eisenhart}.
The higher terms of O($ (\varphi^i)^3$) are unnecessary for our present purpose.
If we take $\varphi^i$ as independent variables but not functions of $x^\mu$,
	eq.\ (\ref{X=Y+}) gives the transformation 
	from the coordinate system $(x^\mu,\varphi^i)$ to $X^I$.
The former is called ``Riemannian coordinate system".
In the far regions off the brane in comparison with
	the scale of the brane curvature,
	this coordinate system may encounter singularities
	and multi-definitions.
It causes, however, no problem, 
	since only small fluctuations are necessary for our purpose.
The higher terms in $\varphi^i$, however, may cause another problem
	in practical calculation of the quantum contributions.
They give rise to the higher terms in $\varphi^i$ in ${\cal L}'_{\rm br}$,
	and hence the higher loop diagrams,
	which we have no systematic way to evaluate.
In fact we do not even know how such large fluctuations contribute 
	to the quantum effects.
Here we restricts to retain the contributions 
	only from the small fluctuations of the brane,
	and neglect the terms of O($(\varphi^i)^3$) and the higher
	in ${\cal L}'_{\rm br}$. 
Accordingly, we are to evaluate only the one-loop diagrams.

Now we substitute (\ref{X=Y+}) into the Lagrangian (\ref{NG'}). 
For our purpose, it is sufficient to retain explicit forms for terms
	up to O$((\varphi^i)^3)$.
Then, we obtain
\begin{eqnarray}&&\hskip-15pt
	{\cal L}'_{\rm br}= {\cal L}'_0 + {\cal L}'_{\varphi}
	+{\rm O}((\varphi^i)^3)
\\&&\hskip-15pt	
	{\cal L}'_0  
	=-\frac{\lambda}{2} \sqrt{-g}
	(g^{\mu\nu}Y^I_{\ ,\mu} Y^J_{\ ,\nu}G _{IJ}-2), 
\\&&\hskip-15pt	
	{\cal L}'_{\varphi}=-\frac{\lambda}{2} \sqrt{-g}
	[ (D\varphi)_ \mu{}^i(D\varphi)^ \mu{}_i +  \varphi^i\varphi^j Z_{ij}]
  \label{Lvarphi}
\end{eqnarray}
with
\begin{eqnarray}&&\hskip-15pt
	(D\varphi)_\mu{}^i=\varphi^i_{,\mu}+ \varphi^k A^i{}_{k\mu},
  \label{Dvarphi}
\\&&\hskip-15pt	
	Z_{ij}= B_{i\mu\nu }B_ j {}^{\mu \nu }
	+C_{ij}
  \label{Zmn}
\end{eqnarray}
where $A_{ij\mu}$ and $B_{i\mu\nu}$ are the normal connection
	and the extrinsic curvature, respectively,
	and $C_{ij}$  	is a particular combination of components of 
	the bulk curvature tensor ${\cal R}_{LIJK}$
	at the brane in the Riemannian coordinate system. 
They are defined by
\begin{eqnarray}
	A_{ij\mu}&=&n_i^{\ I}n_{j\ ;\mu}^{\ J}G_{IJ}, 
  \label{Amnmu}
\\
	B_{i\mu\nu}&=&n_i^{\ I} Y _{\ ;\mu\nu}^J G _{IJ}, 
  \label{Bmmunu}
\\ \hskip-15pt	
	C_{ij}\ 
	&=& g^{\mu\nu}Y^ M_{\ ,\mu}\ n_i^{\ I}\ n_j^{\ J}\ Y^N_{\ ,\nu}\ 
	{\cal R}_{MIJN} (Y^I),
  \label{Cij}
\end{eqnarray}
where $n_ {j;\mu }^J $ is the covariant derivative:
\begin{eqnarray}
	n_ {j\ ;\mu }^{\ D} = n^{ \ D}_{n\ ,\mu} 
	+ n_j^J Y^M_{,\mu}\Gamma^D_{JM},. 
  \label{Dn}
\end{eqnarray}
and the bulk curvature tensor is defined by 
\begin{eqnarray}\hskip-15pt 
	{\cal R}^ L _{\ IJ K }
	\!\!\!&=&\!\!\Gamma^ L _{I K,J}-\Gamma^ L _{IJ, K }
	+\Gamma^L_{TJ}\Gamma^T_{IK}
	-\Gamma^L _{TK}\Gamma^T_{IJ}.
  \label{calR}
\end{eqnarray}
We can see that (\ref{Lvarphi}) is the Lagrangian 
	for the quantum scalar fields $\varphi^m$ on the curved brane 
	interacting with the given external fields 
	$A_{ij\mu}$, $B_{i\mu\nu}$ and $C_{ij}$.

\section{Quantum Effects\label{quantum}}

\begin{figure}
\includegraphics[width=6.0cm]{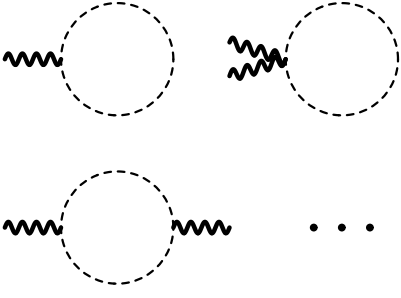}
\caption{The Feynman diagrams.
The dashed lines indicate the $\varphi^i$-propagators,
	and the wavy lines indicate external fields 
	of $A_{ij\mu}$, $B_{i\mu\nu}$ or $h_{\mu\nu}$. 
The dots indicate an infinite series of diagrams 
with a dashed-line loop and with more external wavy lines than two.
By virtue of the symmetries of the system, we have only to calculate
	the three diagrams explicitly drawn here.
}
\label{fig1}
\end{figure}

The quantum effects of the field $\varphi^i$ are described 
	by the effective Lagrangian $ {\cal L}^ {\rm eff }$ 
\begin{eqnarray}&&\hskip-15pt
	\int {\cal L}^{\rm eff}d^4x
	=-i \ln\int [d\varphi^i] \exp \left [i\int {\cal L}'_{\varphi} d^4x\right],
  \label{Seff}
\end{eqnarray}
	where $[d\varphi^i] $ is the path-integration over $\varphi^i $. 
To perform it, we rewrite (\ref{Lvarphi}) into the form \cite{err} 
\begin{eqnarray}
	{\cal L}'_{\varphi}
	= -\frac{\lambda}{2} \varphi_i \left(
	-\delta^i_j \Box + {\cal V}^i{}_j 
	\right) \varphi^j,
  \label{S'V}
\end{eqnarray}
	with $\Box=\eta^{\mu\nu}\partial_\mu\partial_\nu $ and
\begin{eqnarray}
	{\cal V} ^i{}_j
	=\delta^i_j \partial_{\mu} {\cal H} ^{\mu\nu} \partial_{\nu}
	+\partial_{\mu}  {\cal A}^i{}_j{}^\mu 
	+{\cal A}^i{}_j{}^\mu \partial_{\mu}
	+ {\cal Z}^i{}_j
  \label{Vmn}
\end{eqnarray}
\begin{eqnarray}
	{\cal H} ^{\mu\nu}\ \ 
	&=& \eta^{\mu\nu} -\sqrt{-g} g ^{\mu\nu}
  \label{calH=}
\\	
	{\cal A}^i{}_j{}^\mu &=& -\sqrt{-g} A^i{}_j{}^\mu 
  \label{calA=}
\\	
	{\cal Z}^i{}_j{}\hskip5pt &=&
	-\sqrt{-g} (A^{ik\mu} A_{kj\mu}-Z^i{}_j),
  \label{calZ=}
\end{eqnarray}
where the differential operator
	$\partial_\mu\equiv\partial/\partial x ^\mu $ 
	is taken to operate on the whole expression 
	in its right side in (\ref{S'V}). 
The path-integration in (\ref{Seff}) is performed to give 
\begin{eqnarray}
	\int {\cal L}^{\rm eff} d^4 x
	&=&\sum_{n=0}^{\infty}\frac{1}{2ni}{\rm Tr}
	\left(\frac{1}{\Box} {\cal V}^j{}_k \right)^n,
  \label{S'eff1}
\end{eqnarray}
	up to additional constants,
	where Tr indicates the trace over the brane coordinate variable $x^\mu$
	and extra dimension index $j$.
The terms in (\ref{S'eff1}) can be calculated with Feynman-diagram method. 
In terms of the Fourier transforms
\begin{eqnarray}
	\tilde {\cal H}^{\mu\nu}(q_l)
	&=&\int d^4x{\cal H}^{\mu\nu}(x)
	e^{iq_l x} ,
  \label{FourierH}
\\
	\tilde {\cal A}^i{}_j{}^\mu (q_l)
	&=&\int d^4x{\cal A}^j{}_k{}^\mu (x)
	e^{iq_l x} ,
  \label{FourierA}%
\\
	\tilde {\cal Z}^j{}_k(q_l)
	&=&\int d^4x{\cal Z}^j{}_k(x)
	e^{iq_l x} ,
  \label{FourierV}%
\end{eqnarray}
the effective Lagrangian ${\cal L}^{\rm eff}$ is written as
\begin{eqnarray}&&
	{\cal L}^{\rm eff}= \sum_{n=0}^{\infty}\frac{1}{2n}
	\prod_{l=1}^{n}\int\frac{d^4q_l}{(2\pi)^4}e^{-iq_l x} 
	G^n,
  \label{FourierGl}%
\\&&
	G^n
	=\int\frac{d^4p}{i(2\pi)^4}
	\prod_{l=1}^{n}\frac{1}{-p_l^2}
	\tilde {\cal V}^{k_l}{}_{k_{l-1}}(p_l,q_l),
  \label{Gl=}%
\\ &&
	\tilde {\cal V}^k{}_{k'} (p_l,q_l)
	=-\delta^k_{k'}
	(p_l{})_\mu( p_{l-1}){}_\nu \tilde{\cal H} ^{\mu\nu}(q_l) 
\cr&&\ \ \ \ \ \ \ 
	-i (p_l+ p_{l-1})_\mu  \tilde {\cal A}^k{}_{k'}{}^\mu (q_l)
	+ \tilde {\cal Z}^k{}_{k'} (q_l),
  \label{tildeVmn}
\end{eqnarray}
where $p_l=p+q_1+\cdots+q_{l}$ and $k_0=k_n$. 
The function $G^n$ is nothing but the Feynman amplitude for
	the one-loop diagram with $n$ internal lines of $\varphi^j$ and 
	$n$ vertices of $\tilde {\cal V}^k{}_{k'} $ (FIG.\ \ref{fig1}).
Unfortunately, the $p$-dependence of the integrand 
	in (\ref{FourierGl}) with (\ref{Gl=}) indicates that 
	the integration over $p$ diverges at most quartically.
The divergences will be regulated in the next section.
Then, we can perform the integration over $p$ 
	to obtain the function $G^n$. 
The $q_l$'s are replaced by differentiation $i\partial_l$
	of the $l$-th vertex function 
	according to the inverse Fourier transformation in (\ref{FourierGl}).
Collecting all the contributions, 
	which are functions of the fields 
	$g_{\mu\nu}$, $A_{ij\mu}$ and $B_{i\mu\nu}$ and their derivatives,
	we can obtain the expression for 
	the effective Lagrangian ${\cal L}^{\rm eff}$.

\section{Divergences and Regularization \label{cutoff}}

The $p$-dependence of the integrand in (\ref{FourierGl}) with (\ref{Gl=})
	indicates that 
	the integration over $p$ diverges at most quartically.
We expect, however, that fluctuations with smaller wave length
	than the brane thickness are suppressed.
Then, the momenta higher than the inverse of the thickness are cut off.
In order to model the cutoff without violating full symmetry of ${\cal L}'_{\rm br}$,
	we introduce three Pauli-Villers regulators $\Phi_r^m$
	with very large mass $M_r$ ($r$=1,2,3), 
	which are taken equal finally, $M_r\rightarrow\Lambda$,
	following the method of the original paper \cite{Akama78}.
Precisely, it amounts to consider the regularized effective Lagrangian 
\begin{eqnarray}
	{\cal L}^{\rm reg}
	={\cal L}^{\rm eff}+\sum_{r=1}^3 C_r {\cal L}^{\rm eff}_{M_r} 
  \label{Sreg}
\end{eqnarray}
where ${\cal L}^{\rm eff}_{M_r}$ is the effective Lagrangian 
	for the quantum effects from ${\cal L}'_{\Phi_r} $ 
	which is the same as ${\cal L}'_{\varphi}$ except that
	$\varphi^i$ is replaced by the regulator field $\Phi_r^i$ with mass $M_r$,
\begin{eqnarray}&&\hskip-15pt
	\int {\cal L}^{\rm eff}_{M_r}d^4 x
	=-i \ln\int [d\Phi_r^i] \exp \left[i\int {\cal L}'_{\Phi_r} d^4 x\right],
  \label{SeffMj}
\\&&\hskip-15pt
	{\cal L}'_{\Phi_r}= {\cal L}'_{\varphi}|_{\varphi=\Phi_r}+
	\frac{ 1}{2}\lambda M_r^2 \sqrt{-g} \Phi_r^i\Phi_r^j \eta_{ij}.
  \label{S'Mj}
\end{eqnarray}
In (\ref{Sreg}), the coefficients $C_r$ are defined 
	by the coupled algebraic equation
\begin{eqnarray}
	\sum_{r=1}^3 C_r =-1,\ \ \ 
	\sum_{r=1}^3 C_r (M_r)^ 2=\sum_{r=1}^3 C_r (M_r)^ 4=0.
  \label{sumCM2k}
\end{eqnarray}
Note that the added mass term also preserves 
	the full symmetry of ${\cal L}'_{\rm br}$. 
Performing the path integration over $\Phi_r^i $, we have
\begin{eqnarray}&&
	\int {\cal L}^{\rm eff}_{M_r} d^4 x
	=\sum_{n=0}^{\infty}\frac{1}{2ni}{\rm Tr}
	\left(\frac{1}{\Box+M_r^2}{\cal V}_{M_r}{}^k{}_{k'} \right)^n,
  \label{Seffj}
\\&&
	{\cal V}_{M_r}{}^k{}_{k'} ={\cal V}^k{}_{k'}+{\cal F} M_r^2\delta^k_{k'},
  \label{VMjmn}
\\&&
	{\cal F} =1-\sqrt{-g},
  \label{calF=}
\end{eqnarray}
with ${\cal V}^k{}_{k'} $ in (\ref{Vmn}).
In terms of the Fourier transform 
\begin{eqnarray}&&
	\tilde {\cal F} (q_l) 	=\int d^4x{\cal F} (x) 	e^{iq_l x} .
  \label{FourierF}%
\end{eqnarray}
we have
\begin{eqnarray}&&\hskip-20pt
	{\cal L}_{M_r}^{\rm eff}= \sum_{n=0}^{\infty}\frac{1}{2n}
	\prod_{l=1}^{n}\int\frac{d^4q_l}{(2\pi)^4}e^{-iq_l x} 
	G_{M_r}^n,
  \label{FourierGMjl}%
\\&&\hskip-20pt 
	G_{M_r}^n
	=\int\frac{d^4p}{i(2\pi)^4}
	\prod_{l=1}^{n}\frac{1}{-p_l^2+M_r^2}
	\tilde {\cal V}_{M_r}{}^{k_l}{}_{k_{l-1}}(p_l,q_l),
  \label{GMjl=}%
\\ && \hskip-20pt 
	\tilde {\cal V}_{M_r}{}^k{}_{k'} (p_l,q_l)
	=\tilde {\cal V}{}^k{}_{k'} (p_l,q_l)
	+\delta^k_{k'} M_r^2\tilde{\cal F}(q_l), 
  \label{tildeVMjmn}
\end{eqnarray}
with $\tilde {\cal V}{}^k{}_{k'} (p_l,q_l)$ in (\ref{tildeVmn}).
In dimensional regularization, 
	the divergent parts of the Feynman amplitude $G_{M_r}^n$ behaves like
\begin{eqnarray}&&\hskip-15pt
	G_{M_r}^n
	\sim \epsilon^{-1}
	({\cal G}_4M_r^4+{\cal G}_2M_r^2+{\cal G}_0),
  \label{GMjleps}
\end{eqnarray}
	where this is evaluated at the spacetime dimension $4-2\epsilon$,
	and ${\cal G}_{2v}$ are the appropriate coefficient functions.
The singularities at $\epsilon=0$ reflect the divergences in the $p$-integration.
We can see that, 
	when they are summed with the coefficients $C_r$ over $r$ in (\ref{Sreg}),
	they cancel out according to (\ref{sumCM2k}).
Therefore, the $p$-integrations in ${\cal L}^{\rm reg}$ converge.
Any positive power contributions of $M_r$ regular at infinity
	vanish according to (\ref{sumCM2k}).
The function $G_{M_r}^n$ involves logarithmic singularities in $M_r$.
In the equal mass limit $M_r\rightarrow \Lambda$, 
\begin{eqnarray}&&
	\sum_{r=1}^3 C_r M_r^4\ln M_r^{2}\rightarrow -\Lambda^4/2
  \label{CM4ln->}
\\&&
	\sum_{r=1}^3 C_r M_r^2\ln M_r^{2}\rightarrow \Lambda^2/2
  \label{CM2ln->}
\\&&
	\sum_{r=1}^3 C_r \ln M_r^{2}\rightarrow -\ln\Lambda^2
  \label{Cln->}
\end{eqnarray}

\section{Classification of the Terms \label{classification}}

Thus the divergent part ${\cal L}^{\rm div}$ of the regularized effective
	Lagrangian ${\cal L}^{\rm reg}$ consists of the terms 
	which are proportional to $\Lambda^4$, $\Lambda^2$ or $\ln\Lambda^2$,
	and are monomials of ${\cal H}^{\mu\nu}$, ${\cal F}$,
	${\cal A}^i{}_j{}^\mu $, ${\cal Z}^i{}_j$, and their derivatives.
The expressions ${\cal H}^{\mu\nu}$, ${\cal F}$,
	${\cal A}^i{}_j{}^\mu $, and ${\cal Z}^i{}_j$ 
	are written in terms of the fields 
	$g_{\mu\nu}$, $A^i{}_j{}^\mu$, and $Z^i_{\ j} $
	according to (\ref{calH=}),  (\ref{calF=}), (\ref{calA=}), 
	and (\ref{calZ=}).
Introducing the notation 
	$h_{\mu\nu}\equiv g_{\mu\nu}-\eta_{\mu\nu}$,
	we rewrite $ g^{\mu\nu}$ and $\sqrt{-g}$ in ${\cal H}^{\mu\nu}$, ${\cal F}$,
	${\cal A}^i{}_j{}^\mu $, and ${\cal Z}^i{}_j$
	according to
\begin{eqnarray}
	g^{\mu\nu}&=& \eta^{\mu\nu}-h^{\mu\nu}+h_{(2)}^{\mu\nu} 
	+h_{(3)}^{\mu\nu} +\cdots ,
  \label{g^mumu-inh}
\\
	\sqrt{-g}&=& 1+h/2- h_{(2)}/4+ h^2/8+\cdots, 
  \label{sqrt-g-inh}
\end{eqnarray}
with \cite{h^superscripts}
\begin{eqnarray}&&
	h_{(n)}{}^\mu{}_\nu
	=\overbrace{ h^\mu{}_\sigma h^\sigma{} _\tau \cdots h^\rho{} _\nu } ^n,
  \label{h(2)munu}
\\&&
	h = h^\mu{} _\mu,\ \ \ \ \ \ 
	h_{(n)}= h_{(n)}{} ^\mu{} _\mu. 
  \label{h,h(2)}
\end{eqnarray}
Then, ${\cal L}^{\rm div}$ becomes an infinite sum of monomials of 
	$h_{\mu\nu}$, $A^i{}_j{}_\mu$, $Z^i{}_j$,
	and their derivatives.
Let us denote the numbers of $h_{\mu\nu}$, 
	$A^i{}_{j\mu}$, $ Z^i{}_j $, and the differential operators
	in the monomial by $N_h$, 
	$N_A$, $N_Z$ and $N_{\partial}$, respectively. 
The Lagrangian ${\cal L}^{\rm reg}$ should have mass dimension 4,
	while $h_{\mu\nu}$, 
	$A^i{}_{j\mu}$, $ Z^i{}_j $, and the differential operator
	has mass dimension 0, 1, 2, and 1, respectively.
Therefore, the numbers $N_A$, $N_Z$ and $N_{\partial}$ are restricted by
\begin{eqnarray}
	N_A+ 2N_Z+ N_{\partial}\le 4-2k_{\rm div},
  \label{N+N+N<}
\end{eqnarray}
where $ k_{\rm div}=2,1,0$ for $\Lambda^4$, $\Lambda^2$, and $\ln\Lambda^2$
	terms, respectively. 
On the other hand, the number $N_h$ of $h_{\mu\nu}$ is not restricted.
The relation (\ref{N+N+N<}) allows only finite numbers of values of
	$N_A$, $N_Z$ and $N_{\partial}$,
	according to which we can classify the terms of ${\cal L}^{\rm div}$.
Each class involves infinitely many terms for arbitrary values of $N_h$.

They are, however, not all independent, because they are related by 
	high symmetry of the system
	under the general coordinate transformations on the brane
	and SO(4) gauge transformations of the normal space rotation.
Owing to the symmetry of the system,
	only finite number of terms are allowed. 
The general coordinate transformation symmetry requires that 
	the effective Lagrangian density is proportional to $\sqrt{-g}$
	times a sum of invariant forms.
We list the allowed invariant forms in TABLE \ref{tab1},
	where $R=R^\mu{}_\mu$, $ R_{\mu\nu}= R^\lambda {}_{\mu \nu\lambda }$, and
\begin{eqnarray}&&
	R^\kappa _{\ \lambda \mu\nu }
	=
	\gamma^\kappa _{ \lambda\nu,  \mu }-\gamma^\kappa _{\lambda \mu,\nu }
	+\gamma^\kappa_{\rho \mu }\gamma^\rho_{ \lambda\nu }
	-\gamma^\kappa _{\rho \nu }\gamma^\rho _{\lambda \mu },\ \ 
   \label{Rdef}
\\&&
	A_{ij\mu\nu}
	=
	A_{ij \nu,\mu}-A_{ij \mu,\nu}
\cr&&\hskip40pt	
	+A_{ik \mu} A^k_{\ j \nu}-A_{ik \nu } A^k_{\ j\mu },
  \label{Amnmunu}
\end{eqnarray}
	is the field strength of the gauge field $A_{ij\mu }$.
Among the invariants, $R^2$, $R_{\mu\nu} R^{\mu\nu}$, 
	and $R_{\mu\nu\rho\sigma} R^{\mu\nu\rho\sigma }$
	are not all independent, but related by Gauss-Bonnet relation.
\begin{table}
\caption{\label{tab1}Invariant forms}
\begin{ruledtabular}
\begin{tabular}{ccccl}
$k^{\rm div}$&$N_A$&$N_Z$&$N_{\partial}$&invariant forms \\
\hline
2&0&0&0&1\\
\hline
1&0&0&2&$R$\\
 &0&1&0&$Z^i{}_i$ \\
\hline
0&0&0&4&
$R^2$, $R_{\mu\nu} R^{\mu\nu}$, $R_{\mu\nu\rho\sigma} R^{\mu\nu\rho\sigma }$\\
&0&1&2&
$R Z^i{}_i $, $ Z^i{}_{i;\mu}{}^\mu $\\
&0&2&0&
$( Z^i{}_i)^2$, $ Z^{ij} Z_{ij}$\\
&2,3,4&0&$4-N_A$&
$A_{ij\mu\nu} A^{ij\mu\nu}$\\
\end{tabular}
\end{ruledtabular}
\end{table}

\section{Calculation \label{calculation}}

Thus we can calculate the coefficients 
	of the term $\sqrt{-g}$ times the invariant forms
	by calculating the lowest order contributions in $h_{\mu\nu}$.
The lowest contributions to the term with $N_A=N_Z=N_\partial=0$ 
	are O$(h_{\mu\nu})$,
	while those to $N_A=N_Z=0$ and  $ N_\partial\not=0$ 
	are O$((h_{\mu\nu})^2)$,
	because the O$(h_{\mu\nu})$ terms are total derivatives.
Therefore, their lowest terms are in the one- and two-point functions $G^1$ and $G^2$.
The only possible form including $A_{ij\mu}$ is $ A_{ij\mu\nu} A^{ij\mu\nu}$
	and its lowest term is of $O((h_{\mu\nu})^0)$, and it is in $G^2$.
Thus, it is sufficient to calculate $G^1$ and $G^2$ 
	in order to determine full contributions to ${\cal L}^{\rm div}$.

From (\ref{GMjl=}), (\ref{tildeVMjmn}) and (\ref{tildeVmn}), 
	they are given by
\begin{eqnarray}&&\hskip-15pt
	G^1_{M_r}=-N_{\rm ex}\tilde {\cal H}^{\mu\nu}I_{\mu\nu}
	+N_{\rm ex}M_r^2\tilde {\cal F}I+ \tilde {\cal Z}^i{}_i I,
  \label{G^1=}
\\&&\hskip-15pt
	G^2_{M_r}=N_{\rm ex}
	\tilde {\cal H}^{\mu\nu} 
	\tilde {\cal H}^{\lambda\rho} J_{\mu\nu\lambda\rho }
	+2i\tilde {\cal H}^{\mu\nu}\tilde{\cal A}^i{}_i{}^\rho J_{\mu\nu\rho}
\cr&&
	-(N_{\rm ex}M_r^2\tilde {\cal F}+\tilde {\cal Z}^i{}_i)
	\tilde {\cal H}^{\mu\nu}(J_{\mu\nu}-q_\mu q_\nu J)/4
\cr&&
	- \tilde {\cal A}^i{}_j{}^\mu\tilde{\cal A}^j{}_i{}^\nu J_{\mu\nu}
	-2i(M_r^2\tilde {\cal F}\delta^i_j+{\cal Z}^i{}_j)
	\tilde{\cal A}^j{}_i{}^\rho J_{\mu}
\cr&&
	+ (N_{\rm ex}M_r^4\tilde {\cal F}\tilde {\cal F}
	+ M_r^2\tilde {\cal F}{\cal Z}^i{}_i
	+\tilde {\cal Z}^i{}_j \tilde {\cal Z}^j{}_i)J,
  \label{G^2=}
\end{eqnarray}
	where $ N_{\rm ex}=D-4$ is the number of the extra dimensions,
	$q_\mu$ is the momentum flowing in and out through the vertices,
	and
\begin{eqnarray}&&\hskip-20pt 
	I=\int \frac{d^4p}{i(2\pi)^4}
	\frac{1}{[-p^2+M_r^2]},
\\&&\hskip-20pt
	I_{\mu\nu}=\int \frac{d^4p}{i(2\pi)^4}
	\frac{p_\mu p_\nu}{[-p^2+M_r^2]},
\\&&\hskip-20pt
	J=\int \frac{d^4p}{i(2\pi)^4}
	\frac{1}
	{[-(p+q)^2+M_r^2][-p^2+M_r^2]},
\\&&\hskip-20pt
	J_{\mu}=\int \frac{d^4p}{i(2\pi)^4}
	\frac{(2p+q)_\mu }
	{[-(p+q)^2+M_r^2][-p^2+M_r^2]},
\\&&\hskip-20pt
	J_{\mu\nu}=\int \frac{d^4p}{i(2\pi)^4}
	\frac{(2p+q)_\mu (2p+q)_\nu}
	{[-(p+q)^2+M_r^2][-p^2+M_r^2]},
\\&&\hskip-20pt
	J_{\mu\nu\rho}=\int \frac{d^4p}{i(2\pi)^4}
	\frac{(p+q)_\mu p_\nu (2p+q)_\rho }
	{[-(p+q)^2+M_r^2][-p^2+M_r^2]},
\\&&\hskip-20pt 
	J_{\mu\nu\lambda\rho}=\int \frac{d^4p}{i(2\pi)^4}
	\frac{(p+q)_\mu p_\nu p_\lambda (p+q)_\rho }
	{[-(p+q)^2+M_r^2][-p^2+M_r^2]}.
\end{eqnarray}
In the dimensional regularization, for large $M_r^2$, they are calculated to be
\begin{eqnarray}&&\hskip-20pt 
	I= \frac{ M_r^{2-2\epsilon }}{16\pi^2 \epsilon },\ \ \ 
	J= \frac{ M_r^{-2\epsilon }}{16\pi^2 \epsilon },\ \ \ 
  \label{Iresult}
\\&&\hskip-20pt 
	J_{\mu}=0, \ \ \ \  
	J_{\mu\nu\rho}=0,\ \ \ \  
\\&&\hskip-20pt
	I_{\mu\nu}
	=- \frac{ M_r^{2-2\epsilon }}{16\pi^2 \epsilon }
	\eta_{\mu\nu},\ \ \ 
\\&&\hskip-20pt 
	J_{\mu\nu}
	=- \frac{ M_r^{-2\epsilon }}{16\pi^2 \epsilon } 
	\left [
	2M_r^2\eta_{\mu\nu}+\frac{1}{3}( q_\mu q_\nu -q^2\eta _{\mu\nu})
	\right], 
\\&&\hskip-20pt
	J_{\mu\nu\lambda\rho}= \frac{ M_r^{-2\epsilon }}{16\pi^2 \epsilon }
	 \bigg[
	\left(\frac{M_r^4}{8}-\frac{M_r^2q^2}{24}+\frac{q^4}{240}\right)
	S_{\mu\nu\lambda\rho}
\cr&&
	-\left(\frac{M_r^2}{12}-\frac{q^2}{60}\right)
	T_{\mu\nu\lambda\rho}
	+\left(\frac{M_r^2}{6}-\frac{q^2}{40}\right)
	T'_{\mu\nu\lambda\rho}
\cr&& \hskip70pt 
	+\frac{1}{30}q_\mu q_\nu q_\lambda q_\rho 
	\bigg] 
  \label{Jresult}
\\&&\hskip-30pt 
{\rm with}
\cr&&\hskip-25pt 
	S_{\mu\nu\lambda\rho}
	=\eta_{\mu\nu}\eta_{\lambda\rho}
	+\eta_{\mu\lambda }\eta_{\nu \rho} 
	+\eta_{\mu\rho }\eta_{\nu \lambda }, 
\\&&\hskip-25pt 
	T_{\mu\nu\lambda\rho}
	=\eta_{\mu\nu}q_{\lambda}q_{\rho}
	+\eta_{\mu \rho }q_{\lambda }q_{\nu }
	+\eta_{\nu\lambda }q_{\mu }q_ {\rho }
	+\eta_{\lambda\rho }q_{\mu }q_{\nu },
\\&&\hskip-25pt 
	T'_{\mu\nu\lambda\rho}
	=\eta_{\mu\lambda }q_{\nu }q_{\rho}
	+\eta_{\nu\rho }q_{\mu }q_ {\lambda }.
\end{eqnarray}
We substitute (\ref{Iresult})--(\ref{Jresult}) 
	into (\ref{G^1=}) and (\ref{G^2=}),
	and substitute them into (\ref{FourierGl}) to get $ {\cal L}^{\rm eff}$,
	and rearrange the terms into a sum of monomials of 
	$h_{\mu\nu}$, $A^i{}_{j\mu}$, $Z^i{}_j$ and their derivatives.
The divergent terms are 
	proportional to $ M_r^{2v-2\epsilon }/\epsilon $ ($v=0,1,2$). 
When summed over $r=1,2,3$ in the regularization (\ref{Sreg}),
	the singularities of $1/\epsilon $ in ${\cal L}^{\rm reg}$
	cancel out owing to (\ref{sumCM2k}).
Then, in the limit of $\epsilon \rightarrow 0$, 
	there remain the factors $M_r^{2v}\ln M_r^2 $, 
	and, in the equal mass limit $ M_r \rightarrow \Lambda $,
	eqs.\ (\ref{CM4ln->})--(\ref{Cln->}) indicate that
\begin{eqnarray}&& 
	{\sum}_r C_r M_r^{-2\epsilon }/\epsilon \rightarrow \ln\Lambda^2,\ \ 
\\&&
	{\sum}_r C_r M_r^{2-2\epsilon }/\epsilon \rightarrow - \Lambda^2/2,\ \ 
\\&&
	{\sum}_r C_r M_r^{4-2\epsilon }/\epsilon \rightarrow \Lambda^4/2,\ \ 
\end{eqnarray}

The terms are classified as follows.
\\(i) The terms with $N_A=N_Z=0$ are given by \cite{h^superscripts}
\begin{eqnarray}&& 
	\frac{ N_{\rm ex}}{32 (4\pi)^2} \bigg[
	\frac{ \Lambda^4}{2} (4h-2h_{(2)} +h^2)
\cr&&
	+\frac{ \Lambda^2}{3} 
	(h^{\mu\nu,\lambda} h_{\mu\nu,\lambda}
	-2h^{\mu\nu}{}_{,\nu} h_{\mu}{}^{\lambda}{}_{,\lambda}
	+2h^{\mu\nu}{}_{,\nu} h_{,\mu}
	-h^{,\mu} h_{,\mu})
\cr&&
	+\frac{ \ln\Lambda^2}{15} 
	\big \{ h^{\mu\nu,\lambda\rho} h_{\mu\nu,\lambda\rho}
	-2h^{\mu\nu}{}_{,\nu\rho } h_{\mu}{}^{\lambda}{}_{,\lambda\rho }
\cr&&\hskip20pt
	+4(h^{\mu\nu}{}_{,\mu\nu})^2 
	-6h^{\mu\nu}{}_{,\mu\nu} h^{,\lambda}{}_{\lambda} 
	+3(h^{,\mu}{}_{\mu})^2\big\} \bigg] 
  \label{hhh}
\end{eqnarray}
up to total derivatives.
Because the full expression should have the symmetry, 
	they should be the lower order expression of 
	$\sqrt{-g}$ times the invariant forms in table \ref{tab1}.
The terms in (\ref{hhh}) are to be compared with 
	the lower contributions for $\sqrt{-g}$ in (\ref{sqrt-g-inh}) and
\begin{eqnarray}&& 
	\sqrt{-g}R=\frac{1}{4}    
	 (h^{\mu\nu,\lambda} h_{\mu\nu,\lambda}
	-2h^{\mu\nu}{}_{,\nu} h_{\mu}{}^{\lambda}{}_{,\lambda}
\cr&&\hskip80pt
	+2h^{\mu\nu}{}_{,\nu} h_{,\mu}
	-h^{,\mu} h_{,\mu}),
\\&&
	\sqrt{-g}R^2=
	(h^{\mu\nu}{}_{,\mu\nu})^2 
	-2h^{\mu\nu}{}_{,\mu\nu} h^{,\lambda}{}_{\lambda} 
	+(h^{,\mu}{}_{\mu})^2,
\\&&
	\sqrt{-g}R_{\mu\nu}R^{\mu\nu}=-\frac{1}{4} 
	\big [ h^{\mu\nu,\lambda\rho} h_{\mu\nu,\lambda\rho}
	-2h^{\mu\nu}{}_{,\nu\rho } h_{\mu}{}^{\lambda}{}_{,\lambda\rho }
\cr&&\hskip20pt
	+2(h^{\mu\nu}{}_{,\mu\nu})^2 
	-2h^{\mu\nu}{}_{,\mu\nu} h^{,\lambda}{}_{\lambda} 
	+ (h^{,\mu}{}_{\mu})^2\big],
\end{eqnarray}
where total derivatives are neglected.
Note that we have changed the sign convention 
	of the curvature tensor $R^\kappa_{\ \lambda \mu\nu }$ (\ref{Rdef})
	from that of the previous paper \cite{Akama:2013tua}.
\\(ii) The lowest contributions to ${\cal L}^{\rm reg}$ 
	with $N_Z\not=0$ and $N_\partial=0$ are 
\begin{eqnarray}&&\hskip-10mm
	\frac{1}{4(4\pi)^2} \left(
	{\Lambda^2}Z^i{}_i 
	+{\ln\Lambda^2}Z^{ij}Z_{ij}	\right),
\end{eqnarray}
which are taken as the lowest parts of the forms 
	$\sqrt{-g} Z^i{}_i $ and $\sqrt{-g} Z^{ij}Z_{ij}$.
\\(iii) The lowest contribution with $N_Z\not=0$ and $N_\partial=2$ is
\begin{eqnarray}&&\hskip-10mm
	\frac{\ln\Lambda^2}{12(4\pi)^2} 
	( h^{\mu\nu}{}_{,\mu\nu}+ h^{,\mu}{}_{\mu}) Z^i{}_i 
\end{eqnarray}
which is the lowest part of the form 
	$\sqrt{-g}R Z^i{}_i $.
\\(iv) The lowest contribution with $N_A\not=0$ is 
\begin{eqnarray}&&\hskip -10mm
	\frac{\ln\Lambda^2}{24(4\pi)^2} 
	(A_{ij\mu,\nu}-A_{ij\nu,\mu})(A^{ij\mu,\nu}-A^{ij\nu,\mu}),
\end{eqnarray}
which is the lowest part of the form 
	$ \sqrt{-g}A_{ij\mu\nu} A^{ij\mu\nu}$ with $N_A=2$.
Note that it suffices to determine the coefficient of the form in ${\cal L}^{\rm reg}$.

Collecting the results of (i)--(iv), 
	we finally obtain the expression for 
	the divergent part ${\cal L}^{\rm div}$ of ${\cal L}^{\rm reg}$:
\begin{eqnarray}&&\hskip-10pt 
	{\cal L}^{\rm div}\!\!\!=\frac{\sqrt{-g}}{(4\pi)^2}
	\bigg[N_{\rm ex}\left(
	\frac{\Lambda^4}{8}
	+\frac{\Lambda^2}{24}R       
	+\frac{\ln\Lambda^2}{240}(R^2 + 2R_{\mu\nu} R^{\mu\nu})
	\right)
\cr&&\hskip15mm 
	+\frac{ \Lambda^2}{4}Z_i{}^i   
	+\frac{\ln\Lambda^2}{4} Z_{ij}Z^{ij}  
	+\frac{\ln\Lambda^2}{12}R
	Z_i{}^i     
\cr \cr&&\hskip15mm 
	-\frac{\ln\Lambda^2}{24}A_{ij\mu\nu} A^{ij\mu\nu}
	\bigg].
  \label{Ldiv}
\end{eqnarray}
In terms of the fields $A_{ij\mu}$, $B_{i\mu\nu}$ and $C_{ij}$,
(\ref{Ldiv}) is rewritten as
\begin{eqnarray}&&\hskip-10pt 
	{\cal L}^{\rm div}\!\!\!=\frac{\sqrt{-g}}{(4\pi)^2}
	\bigg[N_{\rm ex}\left(
	\frac{\Lambda^4}{8}
	+\frac{\Lambda^2}{24}R       
	+\frac{\ln\Lambda^2}{240}(R^2 + 2R_{\mu\nu} R^{\mu\nu})
	\right)
\cr&&\hskip15mm 
	+\frac{\Lambda^2}{4} B_{i\mu\nu }B^{i\mu \nu } 
	+\frac{\ln\Lambda^2}{4} 
	B_{i\mu\nu }B_ j ^{\ \mu \nu } B^i_{\ \lambda\rho}B^{j\lambda\rho} 
\cr&&\hskip15mm 
	+\frac{\ln\Lambda^2}{12}R B_{i\mu\nu }B^{i\mu \nu }  
	-\frac{\ln\Lambda^2}{24}A_{ij\mu\nu} A^{ij\mu\nu}
\cr&&\hskip15mm 
	+\frac{ \Lambda^2}{4}
	C_i{}^i     
	+\frac{\ln\Lambda^2}{4} 
	C_{ij}C^{ij}   
\cr \cr&&\hskip15mm 
	+\frac{\ln\Lambda^2}{12}R
	C_i{}^i     
	+\frac{\ln\Lambda^2}{2} 
	B_{i\mu\nu }B_j^{\ \mu \nu }
	C^{ij}     
	\bigg],
  \label{Sdiv}
\end{eqnarray}
where	$ A_{ij\mu\nu}$ is the field strength (\ref{Amnmunu}) of the
	normal-connection gauge field $A_{ij\mu}$ (\ref{Amnmu}).
The divergences cannot be renormalized 
	because the original action does not have these terms.
They are, however, cut off by the momentum cut off
	at the inverse of the brane thickness.
The fluctuations with smaller wave length than the brane thickness
	make no sense, and do not contribute to the quantum loop effects.
Therefore, the contributions in (\ref{Sdiv})
	give rise to genuine quantum induced effects.

\section{Conclusions and Discussions\label{conclusion}}

We have established a precise formalism to derive the expressions 
	for the quantum induced effects due to small fluctuations 
	of the braneworld embedded in a curved bulk.
Assuming general coordinate invariance both of the brane and of the bulk, 
	we adopted the simplest model with the Nambu-Goto action
	(\ref{NG}) of the brane.
In the previous paper \cite{Akama:2013tua}, we inquired this problem 
	for the limited case with a flat bulk.
In this paper, we extended it to the fully general case 
	where the bulk is arbitrarily curved.
To define the brane fluctuations in the curved bulk invariantly,
	we introduced the Riemannian coordinate $\varphi^m$ (\ref{X=Y+})
	for the normal geodesic subspace of the braneworld.
Then, we worked out the quantum effects of the small fluctuations
	using the methods developed in the previous paper.
It turned out that we can systematically incorporate the effects of bulk curvature, 
	and that the induced effects depend also 
	on the bulk curvature components at the brane,
	in addition to the extrinsic curvature and the normal-connection gauge field.
The resultant expression for the induced effects 
	is given by (\ref{Sdiv}).

The first three lines in the right hand side of (\ref{Sdiv}) have the same form
	as the result (88) with flat bulk \cite{Akama:2013tua},
while the last two lines explicitly depend on the bulk curvature components.
In the first line, the first term in the big curly bracket
	contributes a huge amount to the cosmological term, 
	and suffers from the notorious problem of naturalness, 
	as was discussed in the previous paper \cite{Akama:2013tua}. 
It requires unnatural fine-tuning among the parameters 
	and the spacetime configurations,
	though it is possible.
If the cosmological term is successfully suppressed in total,
	the terms with the brane curvature in the big curly bracket
	provide 
	the kinetic term of the metric$g_{\mu\nu}$, 
	i.\ e. the gravity is effectively induced. 
For small curvatures, it is dominated by the Einstein-Hilbert action
	(the second term in the big curly bracket)
	with small corrections from the terms quadratic 
	in the curvature 
	(the last term in the big curly bracket).
An approximate Einstein equation takes place effectively.
The sign of the term is right to give ordinary attractive gravity 
	in accordance with the observation.
Its magnitude indicates 
	that the cutoff $\Lambda$,
	and hence, the brane thickness is order of the Planck scale.

The second line in (\ref{Sdiv}) includes the mass term (the first term) 
	of the extrinsic curvature $B_{i\mu\nu}$
	and its self-interaction terms (the second term).
The field $B_{i\mu\nu}$ has no kinetic term
	and a kind of massive background field which does not propagate.
The mass is order of the Planck scale 
	and the self-interactions are suppressed compared with its mass.
The first term in the third line gives rise to 
	an interaction of $B_{i\mu\nu}$ with $R^\kappa{}_{\!\!\mu\nu\lambda}$,
	and hence, with $g_{\mu\nu}$ through it.
The interactions are also suppressed compared with its mass.
The second term in the third line is nothing but the gauge invariant
	kinetic term of the gauge field  $A_{ij\mu}$.
This term includes the kinetic term and the self-interaction terms of $A_{ij\mu}$.
The field $A_{ij\mu}$ looks like a composite gauge field.
This has the gauge symmetry of the rotational group SO$(N_{\rm ex})$
	of the normal space to the braneworld.
The magnitude coupling constant is order of $1/\sqrt{N_{\rm ex}\ln\Lambda^2}$.
This can be a candidate of the origin of existing gauge symmetries
	in the particle theories \cite{Akama87}.
The fields $A_{ij\mu}$ and $B_{i\mu\nu}$ interact also with $g_{\mu\nu}$ 
	through the factors $\sqrt{-g}$ and $g^{\mu\nu}$ 
	which is the inverse of $g_{\mu\nu}$.
In summary, the terms in the second and the third lines describe
	the behaviors of the fields $A_{ij\mu}$ and $B_{i\mu\nu}$
	interacting with the gravitational fields $g_{\mu\nu}$
	induced from the terms in the first line.

Though these terms have the same forms as 
	the result (88) with flat bulk \cite{Akama:2013tua},
	the fields $g_{\mu\nu}$, $A_{ij\mu}$ and $B_{i\mu\nu}$ 
	are related to the bulk curvature ${\cal R}^L_{\ IJK}$ at the brane,
	through the Gauss-Codazzi-Ricci equations:
\begin{eqnarray}&&\hskip-15pt 
	R _{\rho\lambda\mu\nu }
	+B_{i\rho \nu} B^i_{\  \lambda\mu}-B_{i\rho \mu} B^i_{\ \lambda\nu}
	=
	\overline{\cal R} _{\rho \lambda\mu\nu },
   \label{Gauss}
\\&&\hskip-15pt
	B_{i\lambda[\nu,\mu]}
	+A_{ij \mu} B^j_{\ \lambda\nu}- A_{ij \nu} B^j_{\ \lambda\mu}
	=
	\overline{\cal R} _{i\lambda\mu\nu},
   \label{Codazzi}
\\&&\hskip-15pt
	A_{ij\mu\nu }
	+B_{i \nu\rho}B_{j\mu}{}^\rho- B_{i \mu\rho}B_{j\nu}{}^\rho 
	=\overline{\cal R}_{ij\mu\nu},
   \label{Ricci}
\end{eqnarray}   
where 
	$\overline{\cal R}_{\rho\lambda \mu\nu }$, 
	$\overline{\cal R}_{i\lambda \mu\nu }$ and 
	$\overline{\cal R}_{ij\mu\nu}$ 
	are particular components of the bulk curvature tensor
	${\cal R}_{IJKL}$
	at the brane in the Riemannian-coordinate:
\begin{eqnarray}&&\hskip-10pt 
	\overline {\cal R}_{\rho\lambda\mu\nu}
	= Y^I_{\ ,\rho}Y^J_{\ ,\lambda} Y^K_{\ ,\mu} Y^L_{\ ,\nu} 
	{\cal R}_{IJKL} (Y^I),
  \label{Rcomp1}
\\&&\hskip-10pt
	\overline {\cal R}_{i\lambda\mu\nu}
	= n_i^{\ I}Y^J_{\ ,\lambda} Y^K_{\ ,\mu} Y^L_{\ ,\nu} 
	{\cal R}_{IJKL} (Y^I),
  \label{Rcomp2}
\\&&\hskip-10pt
	\overline {\cal R}_{ij\mu\nu}
	= n_i^{\ I} n_j^{\ J} Y^K_{\ ,\mu} Y^L_{\ ,\nu} 
	{\cal R}_{IJKL} (Y^I). 
  \label{Rcomp3}
\end{eqnarray}
If the bulk curvature were determined by some condition in the bulk physics, 
	the configurations of 	
	the fields $g_{\mu\nu}$, $A_{ij\mu}$ and $B_{i\mu\nu}$ 
	should be restricted by the Gauss-Codazzi-Ricci equations
	(\ref{Gauss})--(\ref{Ricci}).
In fact, they are the integrablility condition for the embedding of the brane.
Unfortunately, it may be very difficult to know the 
	physical conditions to determine the bulk curvature.

On the other hand, the last two lines (the forth and the last) in (\ref{Sdiv})
	explicitly depend on the bulk curvature at the braneworld 
	through $C_{ij}$ in (\ref{Cij}),
	which is rewritten as $C_{ij}=g^{\mu\nu} \overline {\cal R}_{\mu ij \nu}$
	with
\begin{eqnarray}&&\hskip-10pt 
	\overline {\cal R}_{\mu ij \nu}
	= Y^L_{\ ,\mu} n_i^{\ I} n_j^{\ J} Y^K_{\ ,\nu} 
	{\cal R}_{LIJK} (Y^I). 
  \label{Rcomp4}
\end{eqnarray}
They are absent in the flat-bulk results \cite{Akama:2013tua}.
Note that the components in (\ref{Rcomp4}) are different 
	from those in (\ref{Rcomp1})--(\ref{Rcomp3}) which appear in
	the Gauss-Codazzi-Ricci equations.
The field $C_{ij}$ is entirely
 independent of the fields
	$g_{\mu\nu}$, $A_{ij\mu}$ and $B_{i\mu\nu}$.
We can see this (independence) also from the fact 
	that, in the definition of the bulk curvature (\ref{calR}),
	the components in  (\ref{Rcomp1})--(\ref{Rcomp3})
	do not include the differentiation of the bulk metric 
	with respect to the normal coordinate, 
	while (\ref{Rcomp4}) does.
The forth line of (\ref{Sdiv}) describes behavior of the field $C_{ij}$
	without kinetic term, 
	indicating that it is a background field without propagation.
The last line describes interactions of the field $C_{ij}$ with 
	the scalar curvature $R$ of the brane and 
	those with the extrinsic curvature $B_{i\mu\nu}$.
The field $C_{ij}$ interacts also with $g_{\mu\nu}$
	through the factor $\sqrt{-g}$.
In summary, the terms in the forth and the last lines describe
	the behaviors of the field $C_{ij}$ 
	interacting with the gravitational fields $g_{\mu\nu}$
	induced from the terms in the first line.

\begin{acknowledgments}

We would like to thank 
Professor G.~R.~Dvali,
Professor G.~Gabadadze,
Professor M.~E.~Shaposhnikov, 
Professor I.~Antoniadis, 
Professor M.~Giovannini,
Professor S.~Randjbar-Daemi, 
Professor R.~Gregory, 
Professor P.~Kanti, 
Professor G.~Gibbons, 
Professor K.~Hashimoto, 
Professor E.~J.~Copeland, 
Professor D.~L.~Wiltshire, 
Professor I.~P.~Neupane,
Professor R.~R.~Volkas, 
Professor A.~Kobakhidze, 
Professor C.~Wetterich, 
Professor M.~Shifman, 
Professor A.~Vainshtein, 
Professor D.~Wands, 
Professor M.~Visser, 
Professor T.~Inami, 
Professor I.~Oda and 
Professor H.~Mukaida 
for invaluable discussions and their kind hospitality extended to us
during our stay in their places.

This work was supported by Grant-in-Aid for Scientific Research,
No.\ 13640297, 17500601, and 22500819
from Japanese Ministry of Education, Culture, Sports, Science and Technology.

\end{acknowledgments}



\begin{thebibliography}{99}
\bibitem{h^superscripts}
Note that $h_{\mu\nu}$ and its derivatives are not brane tensors, 
	and their superscripts are not defined by multiplying $g^{\mu\nu}$.
For convenience of calculations, 
	we define here that their subscripts are raised by $\eta^{\mu\nu}$.
\end{thebibliography}
\end{document}